\documentclass[aps,prl, twocolumn,showpacs]{revtex4}
\usepackage{amsmath}
\usepackage{epsfig}
\usepackage{amssymb}
\usepackage{dcolumn}
\usepackage{graphicx}
\usepackage{dcolumn}

\begin{document}
\date{\today}

\title{Driving defect modes of Bose-Einstein condensates in optical lattices}

\author{Valeriy A. Brazhnyi$^1$} 
  
\author{Vladimir V. Konotop$^{1}$}
 
\author{V\'{\i}ctor M. P\'erez-Garc\'{\i}a$^2$}

\affiliation{$^1$Centro de F\'{\i}sica Te\'{o}rica e Computacional, Universidade de Lisboa, 
Complexo Interdisciplinar, Avenida Profesor Gama Pinto 2, Lisbon 1649-003, Portugal
\\
$^2$Dept. de Matem\'aticas, E.T.S.I.  Industriales, 
U. Castilla-La Mancha, 13071 Ciudad Real, Spain}

\begin{abstract}
 We present  an approximate analytical theory 
and direct numerical computation of defect modes of a 
Bose-Einstein condensate loaded in an optical lattice and subject
 to an additional localized (defect) potential. 
 Some of the modes are found to be remarkably stable and can be driven 
 along the lattice by means of a defect moving following a
  step-like function defined by the period of Josephson oscillations and the 
  macroscopic stability of the atoms.
\end{abstract}


\maketitle

\emph{Introduction.-} Loading and manipulating cold bosonic atoms in optical
lattices is a fascinating, rapidly growing branch of cold atom
physics  \cite{1,Rev2,Morsch}. Recent progresses of experimental investigations of Bose-Einstein condensates (BECs) in optical lattices (OLs)~\cite{Morsch} and in particular the direct observation of a gap soliton~\cite{GapSol} have stimulated studies of the nonlinear dynamics of matter waves in periodic media (see e.g. \cite{Morsch,review} and references therein). OLs provide a tool for changing the effective properties of atomic media allowing for the existence of solitary waves in a condensate with positive scattering length. Moreover, OLs provide also ways for manipulating nonlinear matter waves. For instance, smooth modulations of an OL can be used to control the dynamics of gap solitons~\cite{review,BKK}. On the other hand, strongly localized defects generated by narrow laser beams have been used in experiments to generate rotational motion of condensed atoms~\cite{Ket}  and to study scattering of a soliton on a defect for the sake of understanding  BEC expansion in an effectively one-dimensional (1D) random potential~\cite{Fort}. 

In this Letter we study the combined  effect of an OL and of a defect, whose spatial extension is of the order of the OL period, on the spatial distribution and dynamics of matter waves. Specifically we study the existence and stability of stationary modes localized in the vicinity of the defect, concentrating on the approximate analytical description of defect modes and the numerical simulation of their dynamics, establishing the stability properties of the modes.  We also show that a defect mode can be driven through an OL over hundreds of lattice periods, thus representing an effective tool for matter waves management.

\emph{The model.-} In the quasi 1D limit a BEC is governed by the dimensionless 1D Gross-Pitaevskii (GP) equation~\cite{review}
\begin{equation}
\label{NLS}
i\psi_t = -\psi_{xx}+A \cos(2x)\psi+V_d(x-x_d)\psi + \sigma |\psi|^{2}\psi,
\end{equation}
where  $\sigma=1$ for repulsive and $\sigma=-1$ for attractive interactions among atoms, 
 $A$ is the amplitude of the lattice potential, the lattice constant is chosen to be $\pi$ without restriction of generality
and $V_d(x)= (\eta/\sqrt{2\pi}\ell) \exp{\left(-x^2/2\ell^2\right)},$ is a defect potential 
characterized by its amplitude $\eta$, width $\ell \lesssim \pi$ (i.e. localized
on a distance of the order of one lattice period) and center $x_d$. 
This potential can be experimentally obtained by illuminating the condensate with an orthogonal narrow Gaussian laser beam.

Since the defect is strongly localized we can use the expansion over the orthonormal basis of 
Wannier functions (WFs), which are also localized on about one lattice period~\cite{solid}. 
 To do this we first expand the solution of (\ref{NLS}),  $\psi(x,t)=\sum_{\alpha=0}^{\infty}\int dk \varphi_{\alpha k}(x) f_\alpha (k,t)$, over the set of Bloch functions (BFs) $\varphi_{\alpha k}(x)$ solving the eigenvalue problem $\left(-d^2/dx^2+A\cos(2x)\right)\varphi_{\alpha k}=E_\alpha (k)\varphi_{\alpha k}$ where $\alpha = 0, 1, ...$ stands for the number of the zone, $k$ is the wave vector considered in the first Brillouin zone (BZ), $k\in[-1,1]$, and hereafter integrals with respect to $k$ are over the first BZ.  
Substituting the above expansion in Eq.~(\ref{NLS}) one obtains that the envelope $f_\alpha (k)$ satisfies
\begin{multline}
	\label{eq1}
	i\dot{f}_\alpha(k)-E_\alpha(k)f_\alpha(k) 
	=\sigma\int dx 
	\bar\varphi_{\alpha k}(x)|\psi(x,t)|^2\psi(x,t)
	 \\
	+\sum_{\alpha'}\int dk'
	\int dx\bar{\varphi}_{\alpha k}(x)V_d(x-x_d)\varphi_{\alpha' k'}(x)
	f_{\alpha'} (k'). 
\end{multline}
The integration with respect to $x$ is carried out over the real axis hereafter and to shorten notations we drop the temporal argument of the envelope $f_\alpha$.

Next we use the standard definition of WFs:  
$
w_{\alpha n}(x)=(1/\sqrt{2})\int\varphi_{\alpha k}(x)\exp(-i\pi nk)dk 
$
which implies $
	\varphi_{\alpha k}(x)=(1/\sqrt{2})\sum_{n}w_{\alpha n}(x)\exp(i\pi nk)$
and introduce the matrix element $
	V_{\alpha\alpha'}^{nn'}=(1/2)\int dx V_d(x)w_{\alpha n}(x)w_{\alpha' n'}(x)$ describing probability of transitions between two bands, $\alpha$ and $\alpha^\prime$ and between sites $n$ and $n^\prime$  as well as its Fourier transform $
\hat{V}_{\alpha\alpha'}(k,k')=\sum_{n,n'}V_{\alpha\alpha'}^{nn'}\exp\left(i\pi (k'n'-kn)\right)$. 
Then, the last term on the right hand side of (\ref{eq1}) is rewritten in the form
	$
	\sum_{\alpha^\prime}\int dk'\hat{V}_{\alpha\alpha'}(k,k')f_\alpha(k')\,.
	$

\emph{Approximate description.-} Let us start with the case where the defect is localized at the origin, $x_d=0$, which also coincides with a minimum of the lattice potential, $A<0$.
Matter-wave modes of small amplitude, having width $\lambda$, exceeding a lattice period, i.e. satisfying $\lambda\gg \pi$, are well described in the effective mass approximation (EMA) \cite{review}. In that case, the characteristic scale of $f_\alpha(k)$ in the momentum space is much smaller than the vector of the reciprocal lattice which in our case is equal to $2$, i.e.  $\Delta k\ll 1$,
the latter being the scale of variation of $\hat{V}_{\alpha\alpha'}(k,k')$. In other words, in the EMA the envelope $f_\alpha(k)$ is much more localized than $\hat{V}_{\alpha\alpha'}(k,k')$. 

Let us consider localized modes whose carrier wave wavevector $k_0$ borders the BZ: $k_0=1$. Then the spatial domain where $f_\alpha(k)$ significantly differs from zero is given by $k\in(1-\Delta k, 1+\Delta k)$ allowing us to approximate with exponential accuracy
\begin{eqnarray}
\label{eq4}
\int \hat{V}_{\alpha\alpha'}(k,k')f_{\alpha'}(k')dk'\approx \frac{\hat{V}_{\alpha\alpha'}}{2\pi}\int f_{\alpha'}(k')dk'.
\end{eqnarray}
Hereafter $\hat{V}_{\alpha\alpha'}=2\pi \hat{V}_{\alpha\alpha'}(1,1)$, $\hat{V}_{\alpha}=\hat{V}_{\alpha\alpha}$.
 
In order to simplify the nonlinear term in (\ref{eq1}) we use the properties of the WFs (see e.g. \cite{AKKS}) and rewrite it as follows 
  \begin{eqnarray}
\frac{1}{2}\sum_{n_{1,2,3}}\sum_{\alpha_{1,2,3}} 
	\int dk_1dk_2dk_3\delta_{k+k_1,k_2+k_3} W_{\alpha\alpha_1\alpha_2\alpha_3}^{0n_1n_2n_3}
	\nonumber \\
		\label{nl1}
	\times e^{i\pi (k_2n_2+k_3n_3-k_1n_1 )}\overline{f}_{\alpha_1}(k_1)
	f_{\alpha_2}(k_2) f_{\alpha_3}(k_3)
\end{eqnarray}
where   $
	W_{\alpha\alpha_1\alpha_2\alpha_3}^{nn_1n_2n_3}=\int dx w_{\alpha n}w_{\alpha_1 
	n_1}w_{\alpha_2 n_2}w_{\alpha_3 n_3}$.  

The EMA is justified  when  the effective mass, $M_\alpha=  \left[d^2E_\alpha(k)/dk^2\right]^{-1}$, is of order one and  the respective WFs are localized on a very few lattice periods. These conditions are verified for OLs with amplitudes of order of a few recoil energies (in dimensionless units for $|A|\sim 1$) what is illustrated by examples in Table~\ref{table1}. 

Subject to the mentioned conditions we rewrite (\ref{nl1}) as $
	 \sum_{\alpha_{1,2,3}} W_{\alpha\alpha_1\alpha_2\alpha_3} Q_{\alpha_1\alpha_2\alpha_3}$ where
\begin{eqnarray}
	\label{w1}
	&&W_{\alpha\alpha_1\alpha_2\alpha_3}=\frac{1}{2}\sum_{n_1n_2n_3} (-1)^{n_2+n_3-n_1}
	W_{\alpha\alpha_1\alpha_2\alpha_3}^{0n_1n_2n_3} \,,
\\
\nonumber
  &&Q_{\alpha\alpha_1\alpha_2}= 
	\int dk_1dk_2 
	\overline{f}_{\alpha}(k_1+k_2-k) 	f_{\alpha_1}(k_1) f_{\alpha_2}(k_2).
\end{eqnarray}

\begin{figure}[ht]
\epsfig{file=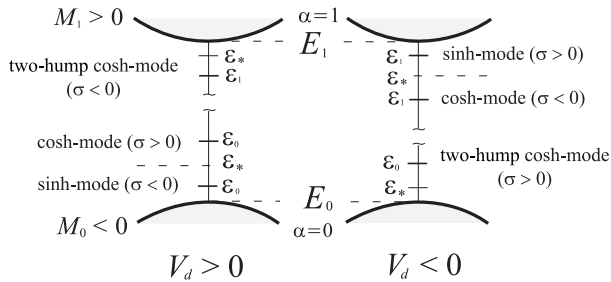, width=8cm}
\caption{Location of the impurity modes for $V_d>0$, $V_d<0$.}
\label{figzero}
\end{figure}

Let us now take into account that for our lattice potential the WFs have the symmetry $w_{\alpha n}(x)=w_{\alpha n}(-x)$ for $\alpha=0,2,...$  and $\quad w_{\alpha n}(x)=-w_{\alpha n}(-x)$ for odd $\alpha$. Then one readily concludes that the even defect $V_d(x)$ results in inter-band transitions only among the bands having the same parity what is reflected by the property $\hat{V}_{\alpha\alpha'}=0$ whenever $|\alpha-\alpha'|$ is odd. On the other hand the nonlinearity, while couples different modes, due to $W_{01}$, does not affect the particle exchange among them preserving the averaged densities.  
Thus, assuming that initially all atoms were loaded in one of the lowest bands one can employ the tight-binding approximation which neglects the inter-band transitions and results in decoupling of the equations for the envelopes $f_\alpha(k)$. Then  Eq.~\eqref{eq1} become
\begin{eqnarray*}
	\label{eqnew}
	i\dot{f}_{\alpha} (k)  =  
  E_{\alpha}(k)f_{\alpha}(k) + \hat{V}_{\alpha}\int f_{{\alpha}} (k)dk   
	 +  \sigma W_{\alpha \alpha} Q_{\alpha \alpha \alpha}\,. 
\end{eqnarray*}
Near the boundary of the BZ we can expand $
E_\alpha(k)\approx E_\alpha+(k-k_{0})^2/2M_\alpha$ where $E_{\alpha}=E_{\alpha}(1)$ is the energy of the $\alpha$-th band at the boundary of the BZ (see Fig.~\ref{figzero}).
Then, introducing $\hat{f}_\alpha(x,t)=e^{-iE_\alpha t}\int dk e^{i(k-k_{0})x}f_\alpha(k,t)$
and after straightforward algebra  we obtain the equation the nonlinear Shr\"{o}dinger (NLS) equation with a delta impurity
\begin{multline}\label{NLS-del}
	i\frac{\partial \hat{f}_{\alpha}}{\partial t}+\frac{1}{2M_{\alpha}} \frac{\partial^2 \hat{f}_{\alpha}}{\partial x^2} 
	 - \sigma W_{{\alpha}{\alpha}} |\hat{f}_{\alpha}|^2\hat{f}_{\alpha}
	= \hat{V}_{{\alpha}}\delta(x)\hat{f}_{\alpha}.
\end{multline}
\begin{table}[ht]
\caption{Parameters of the spectrum at the boundary of the BZ for the two lowest bands. 
The integrals $W_{\alpha\alpha'}$ are approximated with accuracy of about 3\% for $A=-1$ by considering only neighbouring sites in (\ref{w1}). The accuracy of this approximation grows with the OL amplitude.}
		\begin{tabular}{l|ccccccc} \hline\hline
$A$ & $M_0$   & $E_0$ & $W_{00}$ & $M_1$ & $E_1$ & $W_{11}$ & $W_{01}$
\\ \hline
$-1$& $-0.163$ & 0.471 &   0.25   & $0.1$ & $1.467$  & $0.174$ & 0.182 \\
$-5$& $-2.5$  &$-2.076$  &   0.358   &   $0.3$  & $2.5$ &  $0.234$  &0.274
\\ \hline
		\end{tabular}
	\label{table1}
\end{table} 

In Table~\ref{table1} we present the numerical values of the coefficients of (\ref{NLS-del}), used below for the numerical simulations.
 The wave function $\psi(x,t)$ is recovered from Eq. (\ref{NLS-del}) by
\begin{eqnarray}
	\label{recover_fin}
	\psi  \simeq  
	\frac{1}{\sqrt{2}}
	e^{-iE_\alpha t}\sum_n(-1)^n  
	w_{\alpha n}(x)\hat{f}_\alpha (nL).   
\end{eqnarray}

\emph{Defect modes.-} Stationary defect modes of (\ref{NLS-del}) have the form: $\hat{f}_\alpha(x)= e^{-i \varepsilon_\alpha t}\phi_\alpha(x)$ with $\phi_\alpha(x)$  real~\cite{classif}. The applicability of the EMA and the requirement for the frequency of the mode to belong to a forbidden gap, imply the conditions: $|\varepsilon_\alpha|\ll E_{\alpha+1}-E_\alpha$ and $\varepsilon_\alpha M_\alpha<0$ (Fig.~\ref{figzero}).

We discuss below defect modes in the first lowest gap (i.e. for $\alpha=0,1$). 
For $\sigma M_\alpha<0$ one obtains {\em cosh-modes}  
\begin{equation}
\label{cosh}
\phi_\alpha(x)=\frac{\sqrt{2|\varepsilon_\alpha|/ W_{\alpha\alpha}}}{\cosh\left(\sqrt{2|M_\alpha\varepsilon_\alpha|}\left(|x|-x_\alpha\right)\right)}\,.
\end{equation}
Here
$x_\alpha= 
\mbox{atanh} 
\left(\mbox{sign}(M_\alpha\hat{V}_\alpha)\sqrt{\varepsilon_*/\varepsilon_\alpha}\right)/\sqrt{2|M_\alpha\varepsilon_\alpha|},$ and  
$\varepsilon_*=-M_\alpha\hat{V}_\alpha^2/2$.  This mode exists when $|\varepsilon_\alpha|>|\varepsilon_*|$. 
For $M_\alpha\hat{V}_\alpha<0 $ the cosh-mode has only one maximum and otherwise it has a two-hump 
profile ~\cite{classif}.

When $\sigma M_\alpha>0$  there is another localized solution of Eqs. (\ref{NLS-del}) --  a {\em sinh-mode} -- which corresponds to a smaller detuning $|\varepsilon_\alpha|<|\varepsilon_*|$ such that $\varepsilon_\alpha\sigma<0$
\begin{eqnarray}
\label{sinh}
\phi_\alpha(x)=\frac{\sqrt{2|\varepsilon_\alpha|/W_{\alpha\alpha}}}{\sinh\left(\sqrt{2|M_\alpha\varepsilon_\alpha|}(|x|+x_\alpha)\right)}.
\end{eqnarray}
Now 
$x_\alpha=
\mbox{atanh}\left(\sqrt{\varepsilon_\alpha/\varepsilon_*}\right)/\sqrt{2|M_\alpha\varepsilon_\alpha|}$.

Since $w_{1n}$ are odd functions and the defect potential is localized on its zero we have that  $|\hat{V}_{1}|\ll |\hat{V}_{0}|$ ($\hat{V}_{1}=0$ in the limit $\ell=0$). Thus $\hat{f}_0$ is subject to much larger influence of the defect than $\hat{f}_1$. We also notice that since $M_\alpha = \mathcal{O}(1)$ one has  $|\varepsilon_*|\sim \hat{V}_{\alpha}^2$. On the other hand the applicability of our approach implies that all the terms in Eq. (\ref{NLS-del}) are of the same order and that $\lambda\gg\pi$. Thus
$|\varepsilon_\alpha|\sim|\hat{V}_{\alpha}|/\lambda\sim 1/\lambda^2\ll  |E_\alpha|\sim 1$ and $\hat{V}_{\alpha}\ll 1$.  
Below we concentrate on the cosh-mode excited in the vicinity of the lowest band ($\alpha=0$) and with $\sigma=1$. 
\begin{figure}[ht]
\epsfig{file=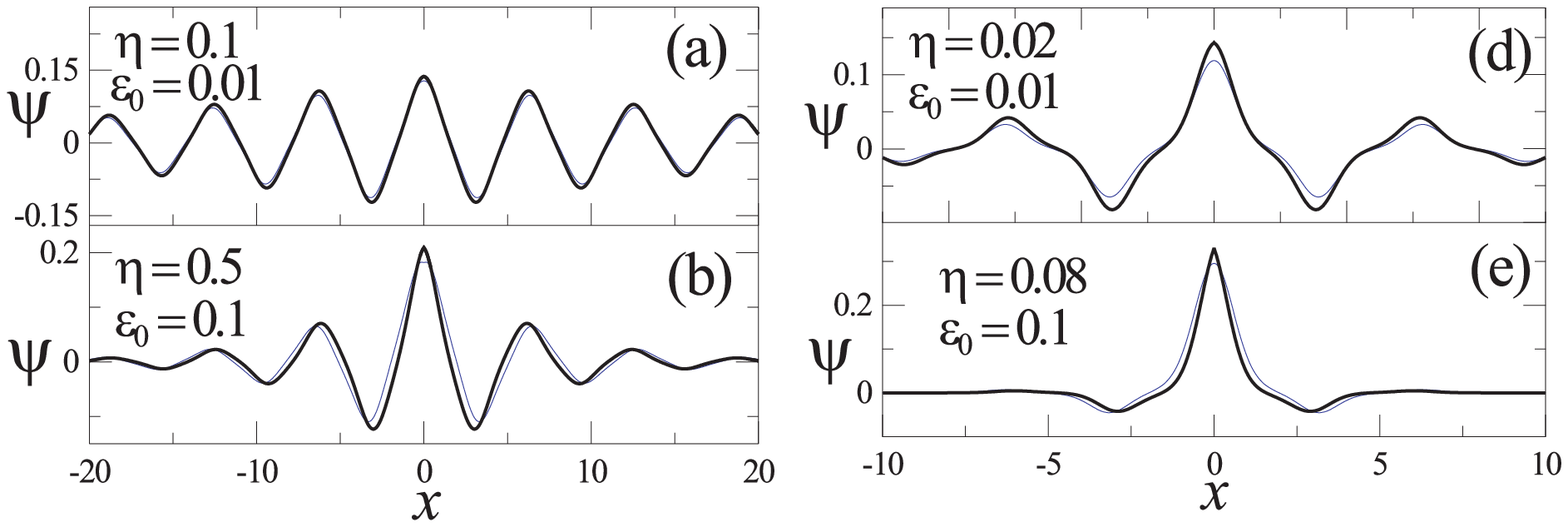,width=\columnwidth}
\epsfig{file=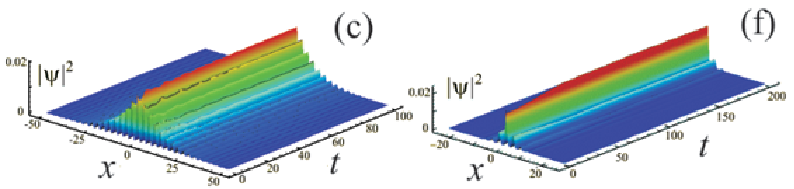,width=\columnwidth}
\caption{Approximate analytical (thick lines) and numerical (thin lines) shapes of the    
stationary cosh-mode of the lowest band   
(a,b) $A=-1$, $\ell=0.1$; 
(d,e) $A=-5$, $\ell=0.1$; 
(c) and (f) dynamics of the modes (a) and (d), correspondingly.
}
\label{figcosh1}
\end{figure}

In Fig.~\ref{figcosh1} we plot both our analytical approximation for cosh-modes given by Eqs.~(\ref{recover_fin}) and (\ref{cosh}) and the direct numerical solutions of Eq.~(\ref{NLS}) for different  OL and defect parameters showing the good accuracy of our one-band approximation.
We have checked that decreasing the defect width by a factor of ten, shapes of the defect modes do not change and their amplitudes differ only by $1\%$, what confirms the excellent accuracy obtained from the delta-approximation for $\ell \ll \pi$. The analytical approximation becomes worse for  increasing   values of the defect strength. 
This occurs because such a defect originates modes localized on a very few lattice periods, and the tight-binding approximation should be used instead~\cite{AKKS}.

\begin{figure}[ht]
\epsfig{file=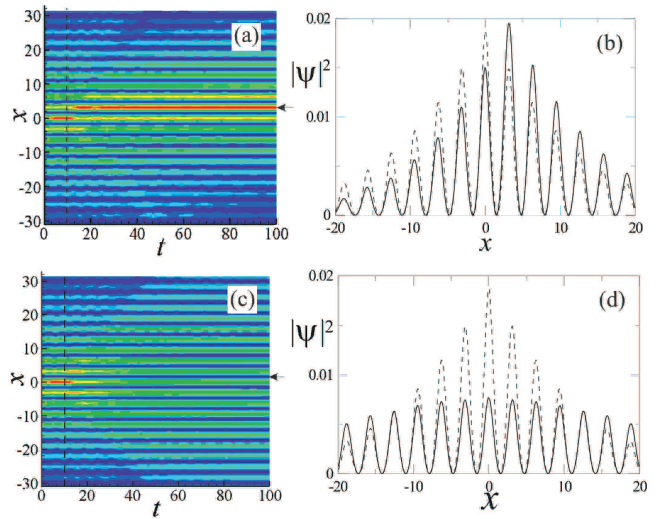,width=8.5cm}
\caption{[Color online] (a-b) Density plots of the dynamics of the cosh-mode with initial parameters  as in  Fig. \ref{figcosh1}(a) and $x_d=0$. At $t=10$ the defect is shifted to the position (a) $x_d=\pi$ and (c) $x_d=\pi/2$ indicated by the arrows. 
In (b) and (d) dashed and solid lines correspond to the initial ($t=0$) and final ($t=100$) profiles of the density of the impurity mode.}
\label{figshift}
\end{figure}

The first simple test of the stability is shown in Fig.~\ref{figcosh1} where we have computed the evolution of the approximate analytical cosh-modes, which introduces a perturbation of the order of a few percents with respect to the true mode. Next we impose stronger perturbations,  by shifting the position of the defect.  First, we shift the defect  at time $t=10$ to the nearest minimum of the periodic potential, $x_d=\pi$, and the mode follows the defect displacement. After some relaxation time the profile of the defect mode centered about $x=\pi$ is very similar to the initial one centered about $x=0$ [Figs.~\ref{figshift}(a-b)].  
When the defect is shifted to $x_d=\pi/2$, the mode is destroyed, since the new position does not correspond to a stable configuration [Figs.~\ref{figshift}(c-d)]. 

\emph{A driven defect mode.-}  The robust mode behavior shown in Fig.~\ref{figshift}(a) suggests that it could be driven by moving the defect along the lattice. On the other hand the mode is destroyed when it  stays close to the local maxima of the potential. This suggests that the defect motion should be defined as a step-like function composed of two characteristic time intervals: a fast one  $\tau$, in which the defect is shifted by one lattice period and a slow one $T$, allowing for the cosh-mode to recover its shape on the new site.

In Fig.~\ref{figshift} (b) we see that the mode dynamics is reflected in the change of its amplitude, what  can be interpreted as a tunneling of a part of atoms in a time $\tau_0$, attracted by the shifted defect.  Since this is a ``single" tunneling process and the intermediate states are unstable the characteristic times should satisfy $\tau\ll \tau_0 \ll T$.

To estimate the Josephson tunneling time $\tau_0$ we take into account that the EMA implies relatively small differences in populations of neighbor potential wells and that we are near the boundary of the BZ, what means that the phase difference between atoms in adjacent cells is $\pi$, thus we can use the results of \cite{RSFS} for the estimation of the half-period of the oscillations of the linear tunneling of a BEC in a double-well potential. This gives $\tau_0\approx 3.67$ for $A=-1$. Hence, the defect must be shifted faster than $\pi/\tau_0\approx 0.86$ and then stop for $T\gg 3.67$ to let the system relax to equilibrium.

\begin{figure}[ht]
\epsfig{file=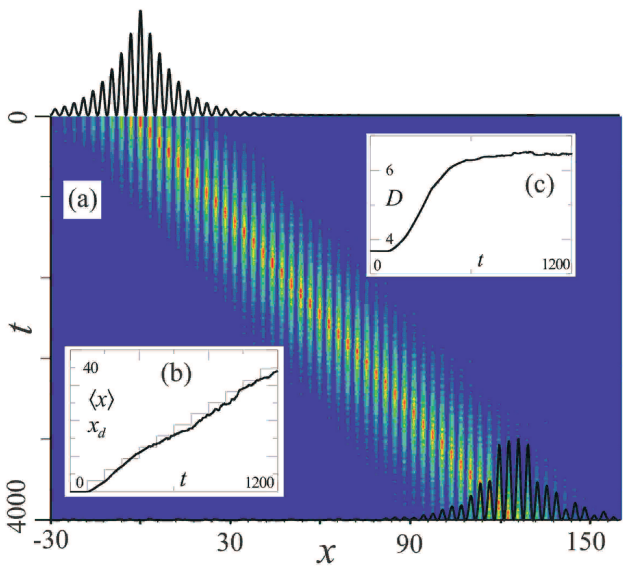,width=8.5cm}
\caption{[Color online]
(a) Pseudocolor and amplitude plots of the motion of the driven defect mode. Initial parameters as in Fig.~\ref{figshift}. (b) Defect position, $x_d(t)$ (thin line), and  average coordinate of the center of the defect mode $\langle x\rangle(t)$ (thick line).  (c) Evolution of the dispersion $D(t)$.
\label{figmove}}
\end{figure}

In Fig.~\ref{figmove} we show numerical simulations of the driving of the numerically found defect modes with $x_d(t)=\pi\sum_{j=0}\theta (t-t_0-100\,j)$, 
where $\theta (t)$ is the Heaviside step function and $t_0$ is a time at which motion is started.  
Shown are also the mean dispersion or the width of the wavepacket $D=\sqrt{\langle x^2\rangle-\langle x\rangle^2}$ where $\langle x^n\rangle=\int x^n|\psi|^2dx/\int |\psi|^2dx$  [Fig.~\ref{figmove}(b-c)].

In Fig.~\ref{figmove}~(c)  we see that the dispersion of the wave packet grows in the interval $t\in(200,600)$ after which the moving mode achieves a stationary profile. This reshaping occurs through the emission of radiation which also contributes to the 
appearance of small oscillations of  $\langle x\rangle$ [see Fig.~\ref{figmove}~(b)] in this time interval.

\begin{figure}[ht]
\epsfig{file=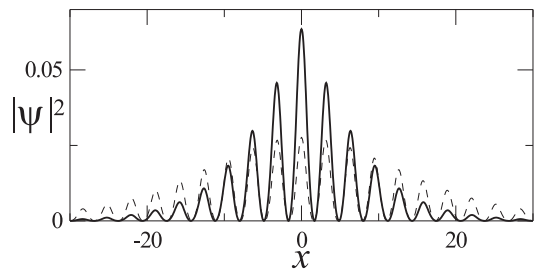,width=7cm}
\caption{
Adiabatic excitation of the cosh-mode. 
Shown are the initial envelope soliton (dashed line) for $A=-1$, $\varepsilon_0=0.01$ and $\eta=0$
and the  final cosh-mode (solid line) corresponding to $\ell=0.1$, $\eta=0.1$,
with $\varepsilon_*=0.0034$ and $\varepsilon_0\approx 0.027$. }
\label{figexcit}
\end{figure}

\emph{Excitation of defect modes.-} A way to excite defect  modes in real situations is to create a gap soliton with energy near the band edge in a homogeneous OL~\cite{GapSol}, 
then by increasing adiabatically the intensity of the transverse laser bean one could generate the defect mode.  We have realized this numerically (see Fig.~\ref{figexcit})
starting from an envelope soliton with energy $E_0+\varepsilon_0$  and increasing $\eta$.
The so obtained final state is the cosh-mode.

\emph{Conclusion.-} We have shown that a BEC  in a 1D OL with an additional  localized potential supports defect modes. In the weak density limit such modes are well described by the effective mass approximation reducing the problem to the NLS equation with a delta impurity. We have constructed the defect modes analytically and numerically and have shown that some of such modes can be driven along the lattice by a defect which follows a specific law of motion. Finally we have shown that defect modes can be easily excited by adiabatically switching on the defect. Our results open new possibilities for controlling matter waves which we hope will stimulate further experiments with Bose-Einstein condensates in optical lattices.

\smallskip

We thank G. L. Alfimov for the software used to obtain Table~I.
V.A.B. was supported by the FCT grant SFRH/BPD/5632/2001. V.A.B. and V.V.K. were supported by the FCT and FEDER under grant POCI/FIS/56237/2004. V. M. P-G. is supported by grants BFM2003-02832 (MEC) and PAC-05-0021 (Junta de Comunidades de Castilla-La Mancha). Cooperative work has been supported by Ac\c{c}\~ao Integrada No E-23/03 and Acciones Integradas  HP2002-0059 (MEC)


\begin{thebibliography}{9}

\bibitem{1}{M. Greiner \emph{et al.}, Nature (London) \textbf{415}, 39 (2002); 
T. Stoferle \emph{et al.}, Phys. Rev. Lett. \textbf{92}, 130403 (2004);
B. Paredes, \emph{et al.}, Nature (London) \textbf{429}, 277 (2004); J. E. Lye \emph{et al.},
Phys. Rev. Lett. \textbf{95}, 070401 (2005).}

\bibitem{Rev2} A. Minguzzi \emph{et al.}, Phys. Rep. \textbf{395}, 223 (2004).

\bibitem{Morsch} O. Morsch and M. Oberthaler, {\it Bose-Einstein condensates in optical lattices}, Rev. Mod. Phys. (to appear).

\bibitem{GapSol} B. Eiermann, \emph{et al.}, Phys. Rev. Lett. {\bf 92}, 230401 (2004).

\bibitem{review} V.A. Brazhnyi and V.V. Konotop, Mod. Phys. Lett. B {\bf 18} 627 (2004). 

\bibitem{BKK}  V.A. Brazhnyi, V.V. Konotop, and V. Kuzmiak, Phys. Rev. A {\bf 70}, 043604  (2004).

\bibitem{Ket}  R. Onofrio \emph{et al.}, Phys. Rev. Lett. {\bf 85}, 2228 (2000).

\bibitem{Fort} C. Fort, \emph{et al.}, cond-mat/0507144.


\bibitem{solid} This technique is well known in solid state physics, see e.g., 
A. I. Anselm, Introduction to Semiconductor Theory (Mir-Moscow, Prentice-Hall, Englewood Cliffs, NJ, 1981).

\bibitem{AKKS} G.A. Alfimov, \emph{et al.},  Phys. Rev. E {\bf 66}, 046608 (2002). 



\bibitem{classif} 
Defecet modes of the NLS equation were considered in A.A. Sukhorukov, \emph{et al.}, Phys. Rev. E {\bf 63}, 036601 (2001).  

 
 

 
 

\bibitem{RSFS} S. Raghavan, \emph{et al.}, Phys. Rev. A, {\bf 59}, 620 (1999).

\end{thebibliography}
\end{document}